\documentclass[10pt,twocolumn,letterpaper]{article}

\usepackage{cvpr}
\usepackage{times}
\usepackage{epsfig}
\usepackage{graphicx}
\usepackage{amsmath}
\usepackage{amssymb}
\usepackage{multirow}
\usepackage{booktabs}
\usepackage{array}
\usepackage{authblk}
\usepackage{footnote}
\usepackage{bm}
\makesavenoteenv{tabular}

\usepackage[breaklinks=true,bookmarks=false]{hyperref}

\cvprfinalcopy 

\def\cvprPaperID{****} 

\begin{document}

\title{Technical report on target classification in SAR track}
\author[1]{Haonan Xu}
\author[1]{Han Yinan}
\author[1]{Haotian Si}
\author[1]{{Yang Yang \thanks{Corresponding author: Yang Yang (yyang@njust.edu.cn)}}}
\affil[1]{Nanjing University of Science and Technology}

\maketitle
\begin{abstract}
This report proposes a robust method for classifying oceanic and atmospheric phenomena using synthetic aperture radar (SAR) imagery. Our proposed method leverages the powerful pre-trained model Swin Transformer v2 Large as the backbone and employs carefully designed data augmentation and exponential moving average during training to enhance the model's generalization capability and stability. In the testing stage, a method called ReAct is utilized to rectify activation values and utilize Energy Score for more accurate measurement of model uncertainty, significantly improving out-of-distribution detection performance. Furthermore, test time augmentation is employed to enhance classification accuracy and prediction stability. Comprehensive experimental results demonstrate that each additional technique significantly improves classification accuracy, confirming their effectiveness in classifying maritime and atmospheric phenomena in SAR imagery.
\end{abstract}

\section{Introduction}
The classification of oceanic and atmospheric phenomena using SAR imagery is a crucial task for maritime safety, environmental monitoring, and climate research \cite{raj2022novel}. Recent advancements in deep learning \cite{yang2021s2osc,yang2021cost,yang2021learning,yang2021corporate,wancovlr,fu2024noise} have facilitated significant progress in image classification tasks \cite{wang2018automated,yang2018complex,yang2019deep,yang2019semi1,yang2019semi}. However, the unique challenges posed by the dynamic and complex nature of oceanic environments require specialized approaches \cite{tao2022classification}. Traditional methods often struggle with the variability and heterogeneity inherent in Synthetic Aperture Radar (SAR) data, leading to the need for more robust and adaptive solutions \cite{li2018sar}.

The goal of this competition is to classify maritime multi-scale remote sensing images and identify various typical marine atmospheric phenomena and targets \cite{li2023remote}. It is also worth noting that our model needs to identify categories in the test data that may be different from those included in the training data, i.e., out-of-distribution (OOD) samples \cite{xiao2021progressive}, This report introduces a simple yet effective method. We carefully designed specialized data augmentation techniques for the dataset and utilized Exponential Moving Average (EMA) to smooth weights during training \cite{singh2022data}. At testing, we improved the model's classification accuracy through Test Time Augmentation (TTA), and enhanced the OOD detection capability by calibrating the penultimate layer activation values with a post-hoc method called ReAct. Finally, we integrated the output results of three models as the result. The combination of these techniques allows for a significant improvement in the classification accuracy of multi-scale oceanic and atmospheric phenomena. The main contributions of this technical report are:

$\bullet$ The carefully designed SAR data augmentation and EMA are applied in the training stage, significantly enhancing the model's generalization ability and robustness.

$\bullet$ During testing, post-hoc methods ReAct are utilized to enhance OOD detection capability. Model ensemble and TTA  techniques are employed to fuse classification results.

$\bullet$ The method we proposed is simple and effective, significantly improving the classification accuracy and the ability to recognize OOD samples. It demonstrates superior performance on the test set.

\section{Methodology}
\begin{figure}[t]
\begin{center}
\includegraphics[width=0.475\textwidth]{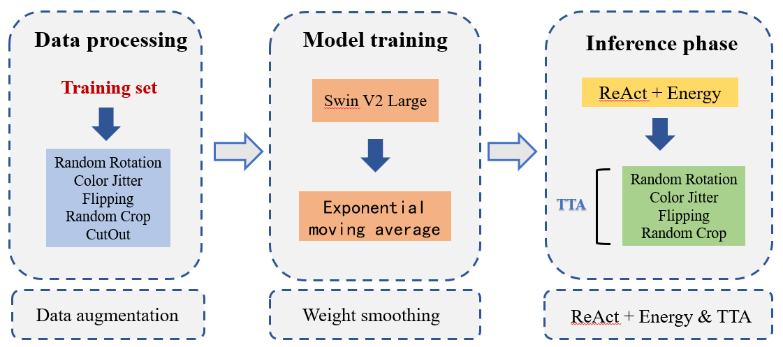}
\end{center}
   \caption{Overall framework.}
   \label{fig:framework}
\end{figure}
In this section, we will provide a detailed description of our proposed method. In Subsection \ref{sec:2.1}, we introduce the details of our model training, and in Subsection \ref{sec:2.2}, we will elaborate on how we enhance model classification accuracy and OOD detection capability during testing. The overall framework diagram is illustrated in Fig. \ref{fig:framework}.

\subsection{Training stage}
\label{sec:2.1}
We utilized the powerful Swin Transformer v2 Large model as the backbone, which has been pre-trained on ImageNet-21k and ImageNet-1k, respectively \cite{conde2022swin2sr}. We fine-tuned the pre-trained models on downstream tasks to effectively improve training efficiency \cite{conde2022swin2sr}. During the training stage, we devised suitable data augmentation and employed the training technique of EMA.

\begin{figure*}
\begin{center}
\includegraphics[width=1.0\textwidth]{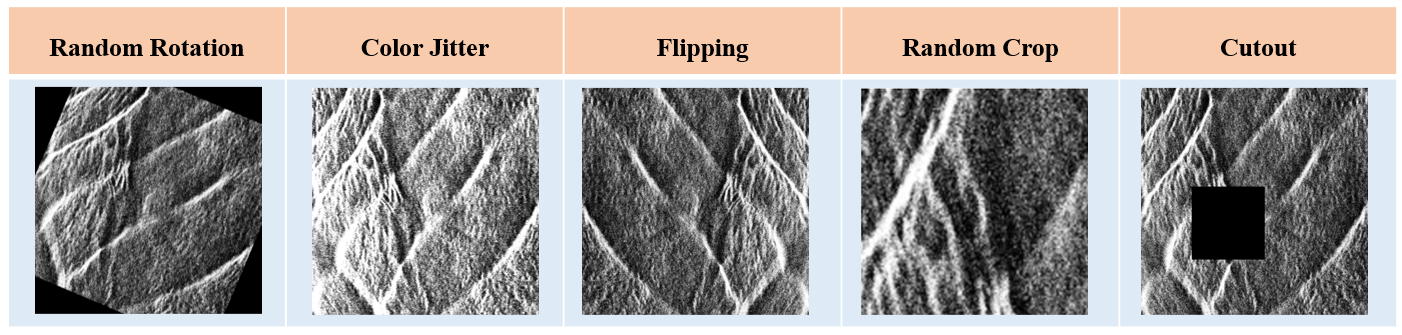}
\end{center}
   \caption{Data augmentation examples. }
   \label{fig:data_aug}
\end{figure*}

\begin{figure*}
\begin{center}
\includegraphics[width=1.0\textwidth]{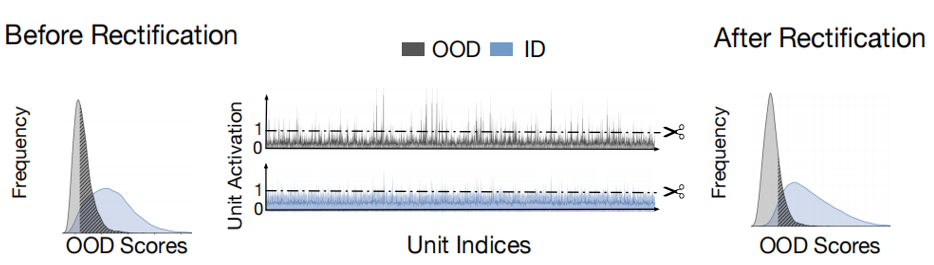}
\end{center}
   \caption{The post-hoc OOD detection method ReAct corrects the activation values of the penultimate layer, increasing the difference between ID data and OOD data, effectively enhancing the OOD detection performance. }
   \label{fig:react}
\end{figure*}

\textbf{Data Augmentation.} We employ a variety of data augmentation techniques to enhance data diversity, prevent model overfitting, and strengthen the model's generalization capability. The example of data augmentation is shown in Fig. \ref{fig:data_aug}.

\textbf{1) Random Rotation}: Images are rotated randomly to make the model less sensitive to object orientation.

\textbf{2) Color Jitter}: Adjustments in brightness and contrast were applied to mimic different lighting conditions and sensor settings.

\textbf{3) Flipping}: Both horizontal and vertical flips were used to enhance the model’s robustness to changes in image orientation.

\textbf{4) Random Crop}: Enhance the diversity of images by randomly cropping them, allowing the model to focus on various parts to aid in classification decisions.

\textbf{5) CutOut}: Parts of the image are randomly obscured to simulate occlusion and encourage focus on various image areas.

\textbf{Exponential Moving Average (EMA).} To enhance model performance, especially in maintaining stability over epochs during training, we integrate the EMA strategy \cite{diqi2023multi}. The EMA approach is utilized to smooth the model parameters, reducing the impact of the high variance in parameter updates that typically occur with stochastic gradient descent \cite{ryder2021neural}. The robustness of the distribution shifts as Eq.(1). For our application, we set the decay rate to 0.99, a common choice that offers a good balance between stability and responsiveness \cite{cai2021exponential}. The model weight update strategy is as follows:
\begin{equation}
\theta_{t}^{ema} = \beta \times \theta_{t - 1}^{ema} + (1 - \beta) \times \theta_{t}^{ema}
\end{equation}
Where $\beta$ is a hyperparameter, and $\theta_{t}^{ema}$ is  the model weight at the epoch.

\subsection{Testing stage}
\label{sec:2.2}
OOD Detection. To manage OOD samples, we used the ReAct \cite{sun2021react} method, which involves clipping features at a predefined threshold before the final classification layer. This step helps in minimizing the influence of anomalous data on the model’s performance. ReAct is to perform post hoc modification to the unit activation, so to bring the overall activation pattern closer to the well-behaved case. Specifically, we consider a pre-trained neural network parameterized by $\theta$, which encodes an input $x~ \in ~R^{d}$ to a feature space with dimension m. We denote by $h(x)~ \in ~R$ the feature vector from the penultimate layer o-f the network. A weight matrix $W~ \in ~R^{m \times k}$ connects the feature $h(x)$ to the output $f(x)$, where $K$ is the total number of classes in $Y~ = ~\left\{ 1,~2,~...,~K \right\}$. The ReAct operation, which is applied on the penultimate layer of a network: 
\begin{equation}
\overset{-}{h}(x) = ReAct\left( h(x);c \right)
\end{equation}
Where $ReAct(x;~c)~ = ~min(x,~c)$ and is applied element-wise to the feature vector $h(x)$. In effect, this operation truncates activations above $c$ to limit the effect of noise. The model output after rectified activation is given by: 
\begin{equation}
f^{ReAct}\left( {x;\theta} \right) = {~W}^{T}\overset{-}{h}(x) + b
\end{equation}
Where $b~ \in ~R^{k}$ is the bias vector. A higher $c$ indicates a larger threshold of activation truncation. When $c~ = ~\infty$, the output becomes equivalent to the original output $f(x;~\theta)$ without rectification, where $
f(x;~\theta)~ = ~W \cdot h(x)~ + ~b$. Ideally, the rectification parameter c should be chosen to sufficiently preserve the activations for ID data while rectifying that of OOD data. In practice, we set $c$ based on the $p$-th percentile of activations estimated on the ID data. For example, when $p = 90$, it indicates that 90\% percent of the ID activations are less than the threshold $c$. The schematic diagram of ReAct is shown in Fig. \ref{fig:react}.
We use the Energy Score to map the model output to a scalar, which is used to measure the model's uncertainty about the output. The energy function is represented as:
\begin{equation}
Energy\left( \mathbf{x} \right) = log\Sigma_{i}exp\left( x_{i} \right)~
\end{equation}
Where $x$ donate the output logits. We use the negative Energy Score as the score function $S(x)$ for OOD detection, and the decision function $G$ can be defined as follows:
\begin{equation}
G(x) = \left\{ \begin{matrix}
{ID~~~~~~if~S(x) > \tau} \\
{OOD~~if~S(x) \leq \tau}
\end{matrix} \right.
\end{equation}
Where $\tau$ is the threshold value. Values greater than t are considered as n-distribution (ID) samples, and those less than $\tau$ are OOD samples. The value of $\tau$ is selected such that 99\% of the training samples are correctly classified.

\begin{figure*}
\begin{center}
\includegraphics[width=1.0\textwidth]{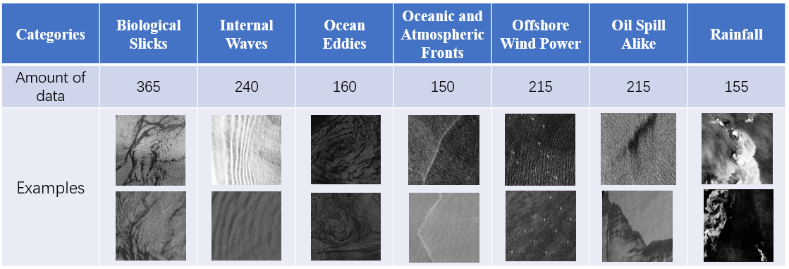}
\end{center}
   \caption{Overview of the training dataset. }
   \label{fig:dataset}
\end{figure*}

\section{Results}
This section presents the experimental results obtained from the application of the Swin Transformer Large model to SAR imagery classification, detailed analyses through ablation studies, and comparative experiments against baseline models. Each experiment is designed to validate the efficacy of the proposed methodologies and to benchmark their performance in real-world scenarios. Experimental Data and Platform. 
\subsection{Dataset} The dataset consists of 1500 SAR images as the training set and 1000 images as the test set. These images contain a wide range of marine and atmospheric phenomena and targets, providing a comprehensive foundation for evaluating the effectiveness of our image classification model. Based on the types of phenomena or objects present, the training set is divided into seven different categories: Biological Slicks, Internal Waves, Ocean Eddies, Oceanic and Atmospheric Fronts, Offshore Wind Power, Oil Spill Alike, Rainfall. The test set contains the same seven categories as the training set and includes some OOD samples. The specific examples for each category are shown in Fig. \ref{fig:dataset}.

\subsection{Implementation Detail}
In our experiment, we fine-tuned the pre-trained weights of the Swin Transformer v2 large provided by Microsoft. We used the Adam optimizer for optimization, set the initial learning rate as 1e-5, and reduced the learning rate by half every 3 epochs if there was no decrease in loss. In total, we trained for 50 epochs. Our experiments are implemented by PyTorch and runs on RTX-4090.

\subsection{Main Results}
The progressive improvement in classification accuracy with the addition of various techniques to the base model is shown in Tab. \ref{table:1}. Each component added enhances the model's performance, reflecting the effectiveness of these methods in improving accuracy for SAR imagery classification tasks.

\begin{table}[!h]
\caption{Impact of Incremental Enhancements on Model Accuracy (Note: The data presented in this table are based on measurements obtained from the online dataset.)}
\label{table:1}
\begin{center}
\begin{tabular}{ >{\centering\arraybackslash}m{5.0cm} 
>{\centering\arraybackslash}m{2.1cm} 
}
 \hline
 Components & Accuracy (\%) \\ 
 \hline
 Base Model & 96.40\% \\
 + Data Augmentation & 97.00\% \\
 + EMA & 97.30\% \\
 + ReAct Method \& Energy Score & 98.70\% \\
 + TTA & 99.20\% \\
\hline
\end{tabular}
\end{center}
\end{table}

\subsection{Ablation study}
\textbf{Effects of EMA.} The decay value controls the rate at which older updates are weighted in the EMA. A higher decay value (closer to 1) results in slower adaptation but smoother overall parameter changes. The selected decay value of 0.99 offers the best trade-off between stability and responsiveness, leading to the most effective enhancement of model performance. Tab. \ref{table:2} displays the effects of varying the decay parameter in the EMA method on the classification accuracy of the model. It emphasizes the optimal balance achieved at a decay value of 0.99, which resulted in the highest classification accuracy.

\begin{table}[!h]
\caption{Impact of EMA Decay Values on Classification Accuracy (Note: The data presented in this table are based on measurements obtained from the training dataset)}
\label{table:2}
\begin{center}
\begin{tabular}{ >{\centering\arraybackslash}m{5.0cm} 
>{\centering\arraybackslash}m{2.1cm} 
}
 \hline
 EMA Decay Value & Accuracy (\%) \\ 
 \hline
 0.90 & 98.90\% \\
 0.95 & 99.00\% \\
 0.99 & 99.20\% \\
 0.995 & 99.10\% \\
\hline
\end{tabular}
\end{center}
\end{table}

\textbf{Effects of TTA}
Justification of 32 TTA Iterations: The number of 32 TTA iterations was chosen based on empirical testing where different iteration counts were evaluated for their impact on classification accuracy and computational efficiency. A balance was struck at 32 iterations, providing a significant improvement in accuracy without excessive computational demand. The accuracy tended to plateau beyond this point, indicating diminishing returns on further increases in iteration count.

\begin{table}[!h]
\caption{Effect of TTA Iterations on Classification Accuracy and Computational Efficiency (Note: The data presented in this table are based on measurements obtained from the training dataset)}
\label{table:3}
\begin{center}
\begin{tabular}{ >{\centering\arraybackslash}m{2.1cm} 
>{\centering\arraybackslash}m{2.1cm} 
>{\centering\arraybackslash}m{3.1cm} 
}
 \hline
 TTA Iterations & Accuracy (\%) & Computational Efficiency (Relative)\\ 
 \hline
 16 & 98.90\% & High \\
 32 & 99.20\% & Moderate\\
 64 & 99.20\% & Low\\
\hline
\end{tabular}
\end{center}
\end{table}

\textbf{Effects of ReAct}
The optimal threshold for the ReAct method was determined by integrating recommendations from the literature [3] and through empirical testing. This combined approach ensured that the selected threshold not only aligns with theoretical best practices but is also practically validated to enhance our model’s robustness and accuracy in handling OOD data. This threshold setting ensures that the model is not only accurate but also stable and reliable when faced with data that differ significantly from the training set.

Tab. \ref{table:4} offer a clear view of how the ReAct method contributes to enhancing model robustness against OOD data, thereby supporting dependable performance across diverse operational environments.

\begin{table}[!h]
\caption{Impact of ReAct Method on Model Robustness and Accuracy (Note: The data presented in this table are based on measurements obtained from the training dataset)}
\label{table:4}
\begin{center}
\begin{tabular}{ >{\centering\arraybackslash}m{3.0cm} 
>{\centering\arraybackslash}m{1.7cm} 
>{\centering\arraybackslash}m{2.3cm} 
}
 \hline
 Configuration & Accuracy (\%) & OOD Handling Effectiveness\\ 
 \hline
 Without ReAct & 99.00\% & Low \\
 With ReAct ($p$ = 0.6) & 99.00\% & Moderate\\
 With ReAct ($p$ = 0.8) & 99.10\% & Moderate\\
 With ReAct ($p$ = 0.9) & 99.20\% & High\\
\hline
\end{tabular}
\end{center}
\end{table}

\section{Conclusion and Applications}
In this report, we presented a robust method for classifying marine and atmospheric phenomena in SAR imagery. Through the application of the Swin Transformer v2 Large model, specialized data augmentation strategies, Exponential Moving Average, ReAct, and Test Time Augmentation, we have significantly improved classification accuracy and the ability to manage Out-of-Distribution samples. Experimental results confirm the effectiveness of these techniques in classifying SAR imagery.
Our model's application is vast. Besides its considerable implications for maritime safety by enabling effective identification of hydrological features and possible hazards, it also opens avenues for critical environmental and climate research. By enhancing the accuracy and generalization capability of the solution, we can effectively monitor oceanic and atmospheric phenomena that play a central role in understanding climate patterns and predicting changes.
In conclusion, we believe our work can contribute significantly to the field of SAR imagery classification, paving the way for more effective applications in maritime safety, environmental stewardship, and climate research.

{\small
\bibliographystyle{ieee}
\bibliography{egpaper_final}
}

\end{document}